\documentclass[12pt]{article}
\usepackage{times}
\usepackage{amsmath}
\usepackage{amssymb}
\usepackage[pdftex]{graphicx} 
\usepackage{epstopdf}         
\usepackage{ITC}

\usepackage{ifpdf}
\ifpdf
\usepackage[pdftex]{graphicx} 
\usepackage{epstopdf}         
\else
\usepackage{graphicx}         
\fi

\title{\Large \vspace{-0.5in}
\MakeUppercase{Spectrally Efficient LDPC Codes for IRIG-106 Waveforms via Random Puncturing}}

    \author{
    Andrew D. Cummins \& David G. M. Mitchell\\
    \normalsize Klipsch School of Electrical and Computer Engineering\\
    \normalsize New Mexico State University, Las Cruces, NM\\
    \normalsize \texttt {\{andrewdc, dgmm\}@nmsu.edu}
    \and 
    Erik Perrins   \\
    \normalsize Department of Electrical Engineering \& Computer Science\\
    \normalsize University of Kansas, Lawrence, KS\\
    \normalsize \texttt {esp@ku.edu}\\
    }

\date{}

\begin{document}

\maketitle 

\section{\MakeUppercase{Abstract}}

\noindent
Low-density parity-check (LDPC) codes form part of the IRIG-106 standard and have been successfully deployed for the Telemetry Group version of shaped-offset quadrature phase shift keying (SOQPSK-TG) modulation. Recently, LDPC code solutions have been proposed and optimized for continuous phase modulations (CPMs), including the pulse code modulation/frequency modulation (PCM/FM) and the multi-h CPM developed by the Advanced Range TeleMetry program (ARTM CPM). These codes were shown to perform around one dB from the respective channel capacities of these modulations. In this paper, we consider the effect of random puncturing of these LDPC codes to further improve spectrum efficiency. We present numerical simulation results that affirm the robust decoding performance promised by LDPC codes designed for ARTM CPM.

\section{\MakeUppercase{INTRODUCTION}}\label{sec:intro}

\noindent
Low density parity check (LDPC) codes \cite{GAL} are a family of forward error correction (FEC) block codes that achieve capacity-approaching performance  under low-complexity iterative decoding schemes \cite{MAC,RICH1}. Efficient LDPC encoder designs have given rise to implementations \cite{LI, RICH2} that meet the stringent requirements of low-power transmitters and LDPC decoding algorithms are well suited to highly-parallelized decoder structures that facilitate high-throughput FEC \cite{HU, CHENG}. LDPC codes have been successfully deployed for aeronautical telemetry systems that employ continuous phase modulations (CPM), specifically SOQPSK-TG \cite{IRIG}. Recent work outlining techniques for the construction of LDPC codes for the remaining two CPM modulation schemes, PCM/FM and ARTM CPM, have demonstrated performance within one dB of channel capacity in numerical simulations \cite{Perrins1, Perrins2}. 

One remaining challenge is to develop spectrally efficient LDPC coding schemes for CPM that can adapt to the changing conditions of time-varying channels.  Coding schemes that can adaptively respond to changing channel conditions without need for alterations to either their encoder or decoder hardware configurations are known as \textit{rate-compatible} codes \cite{HAG}. An efficient means for accomplishing this is to omit sending a portion of the encoded information stream, a process known as \textit{puncturing} 
\cite{HAG, MCL}. This technique  relies on the strength of the code and effect of the puncturing pattern on the decoding algorithm to provide sufficient, though diminished, error correction performance without requiring changes to the decoder \cite{Mitchell1, Mitchell2}. 

In this paper we use random puncturing to explore the trade-off between coding rate and error control performance for some of the LDPC coding schemes proposed in \cite{Perrins1} and \cite{Perrins2}, enabling the development of spectrally-efficient, rate-compatible coding strategies which require minimal alterations to proven hardware implementations. We present numerical simulation results for various puncturing ratios to create effective coding rates in-between those previously demonstrated. To isolate the effect of modulation on the LDPC encoder-decoder pair, we present results both with and without CPM, highlighting the significant coding gain achieved through optimized design of the LDPC codes with respect to specific modulation schemes. We observe that modulation feedback is responsible for greater than $2$ dB of coding gain over the same code without modulation, illustrating the impact of the  optimization for that regime. Our results demonstrate that significant gains in spectral efficiency can be obtained for reasonable degradation in FEC performance.  For example, with $5\%$ of the symbols omitted from transmission by puncturing, we are able to increase a code with rate $R=2/3=0.667$ to $R=0.701$ with less than a $0.2$\,dB loss at a bit error rate of $10^{-6}$ and with no alterations to the component code or decoder required.

\section{\MakeUppercase{BACKGROUND}}\label{sec:background}

\noindent
The transmitter model utilized in our system in shown in Figure 1, with function blocks representing the LDPC encoder, interleaver ($\Pi$), puncturing ($\phi$), CPM modulator, and a module that allows for the insertion of a known symbol sequence referred to as an attached sync marker (ASM) which helps to identify the beginning of each codeword $\{C_n\}$ at the receiver. LDPC encoding is performed by multiplying source information sequence $\textbf{x}$ of length $K$ with the code generator matrix $\textbf{G}$ of size $K \times N$ to form codeword $\textbf{y} = \textbf{x}\textbf{G}$.  

\begin{figure}[t]
\centering
\includegraphics[width = 4.5in]{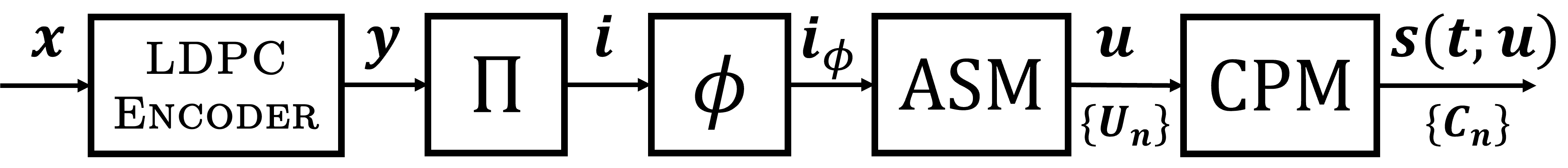}
    
\caption{Model transmitter for punctured LDPC code with CPM modulation.}\label{transmitter}
\end{figure}

To transmit, source information sequence $\textbf{x}$ is encoded blockwise and the resulting codeword $\textbf{y}$ is interleaved to form the sequence $\textbf{i}$, on which puncturing is performed. The resulting punctured, codeword $\textbf{i}_{\phi}$, is concatenated with the ASM to form a frame $\textbf{u}$ which is modulated by the CPM and sent over the communication channel as the sequence of $\textit{q}$-ary frames $s(t;\textbf{u})$ where $q$ is the modulation order. 

Puncturing ($\phi$) must be performed at the encoder prior to the concatenation of the interleaved codeword and ASM in order to avoid desynchronization during decoding. For random puncturing, if the puncturing pattern and interleaver are sufficiently random, puncturing may be performed before or after interleaving as the distribution of punctured symbols will remain uniformly distributed in either case, while nonrandom puncturing would be constrained to placement after interleaving. 

Efficient LDPC encoding algorithms are suitable for many low-power, memory-constrained applications \cite{CHENG}. This is true for the ARTM LDPC codes we have used, which have a quasi-cyclic (QC) structure  in which the code's parity check matrix is an array of sparse repeating units called \textit{circulants} \cite{LI}. The most commonly implemented decoder for LDPC codes is a special case of a more general belief propagation (BP) algorithm 
used for inference on graphical models, known as the sum-product algorithm (SPA) \cite{FRANK}. In this paper, we have used SPA decoding because it provides good performance with soft-input soft-output (SISO) functionality. Efficient hardware implementations of such decoders are numerous and widely adopted \cite{HU}, including reduced complexity variants such as the min-sum algorithm \cite{CHENG}.

The receiver model is shown in Figure 2, with the received sequence $r(t)$ serving as input to the CPM demodulator with no initial \textit{a priori} information ($u=0$). After demodulation of a codeword, it is decoded iteratively in a global loop by a CPM and LDPC decoder which exchange \textit{a priori} information through an interleaver, $\Pi$, identical to that used for encoding, and deinterleaver, $\Pi^{-1}$, its inverse. For the first global iteration the results of the CPM demodulator are permuted and used as inputs to the local LDPC decoder loop by placing the switch in position A.  For subsequent global iterations, the signal flow is altered by placing the switch position in B, with the results of the CPM SISO decoder's updates now serving as inputs to the LDPC decoder.

\begin{figure}[t]
\centering
\includegraphics[width = 6.3in]{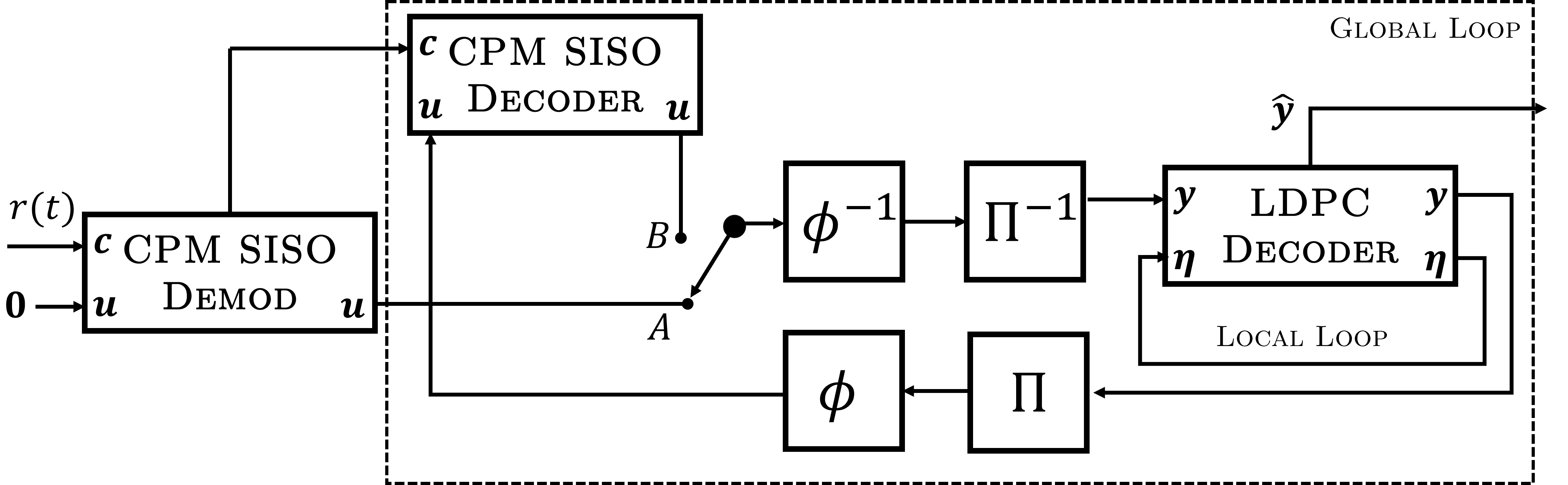}
    
\caption{Model receiver for punctured LDPC code with CPM modulation and iterative decoding.}\label{receiver}
\end{figure}

LDPC SPA decoding is performed using the code's \textit{parity check matrix} $\textbf{H}$, a sparse matrix which satisfies the condition $\textbf{G}\textbf{H}^T = 0$. Sparsity keeps the memory and computational requirements tractable for coding schemes with long block lengths that are needed to obtain capacity-approaching performance. The rows of $\textbf{H}$ represent the parity check conditions of the code while each column represents a codeword symbol.  If a one is present in the $k$-th column of row $j$, code symbol $k$ is found in parity check equation $A_j$.\footnote{An LDPC code with a parity check matrix of constant row and column weight is known as a \textit{regular} code.} 
The LDPC decoder performs iterative SPA decoding in the local loop, before passing the results to the CPM SISO decoder to start the next global iteration and so on. The process continues until some stopping criteria is reached (such a ceiling on the allowed number of iterations) and the resulting estimated codeword $\hat{\textbf{y}}$ is passed out of the decoder. Upon completion, hard decisions on $\hat{\textbf{y}}$ are made to complete decoding. 

Puncturing is performed in the global loop between the CPM SISO decoder and the LDPC decoder.  The module $\phi$ removes codeword soft information symbols in accordance with the known puncturing pattern set by the transmitter, resulting in a truncated sequence $\textbf{u}_{\phi}$, as shown in Figure 3. The module ${\phi}^{-1}$ intercalates zeros into these same puncturing positions to expand the sequence to its original length before interleaving, in a process referred to as \textit{depuncturing}. We note that in our implementation, the LDPC decoder retains soft values on the punctured symbols within the local loop.

\begin{figure}[t]
\centering
\includegraphics[width = 3.2in]{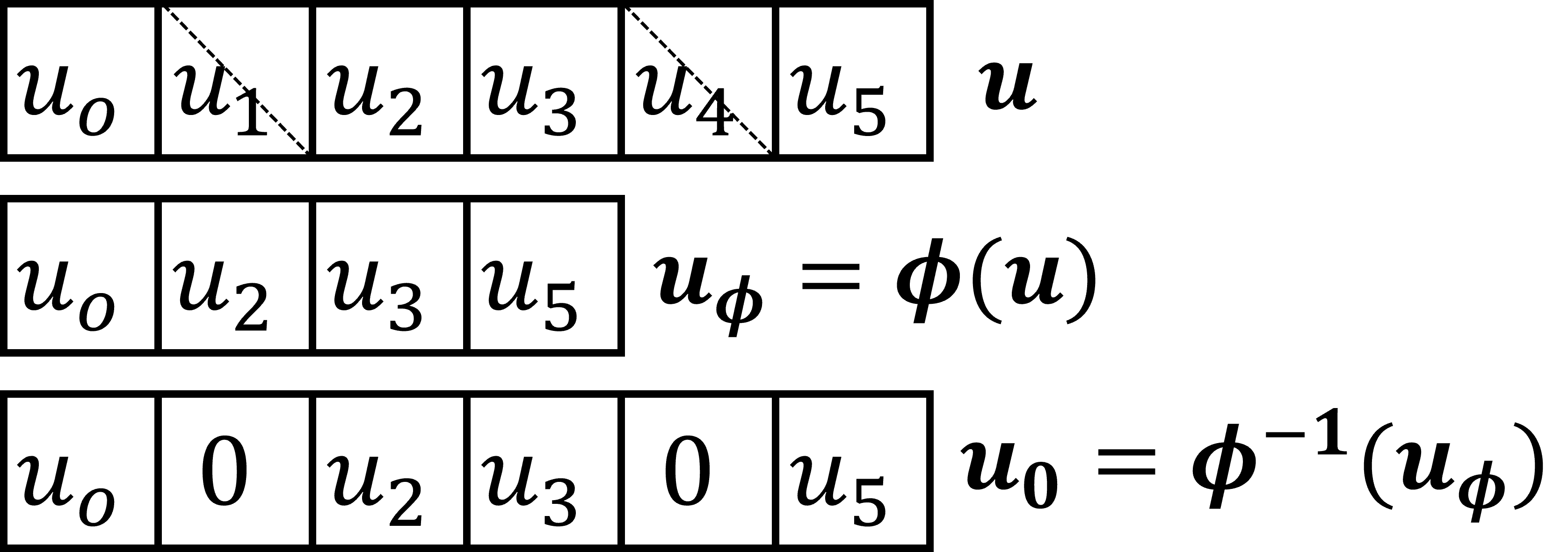}
    
\caption{Puncturing by truncation and depuncturing by zero-padding within the global decoding loop of the receiver.}\label{punct}
\end{figure}

\section{\MakeUppercase{RANDOM PUNCTURING}}\label{sec:punct}

\noindent
The code rate, $R=\frac{K}{N}$ is the ratio of the number of symbols in the source information sequence, $K$, and the number of symbols in the transmitted information sequence, $N$, including parity information.  For a block code of length $N$, wherein $N_{\phi}$ symbols are punctured, we express the amount of puncturing \textit{overhead} as the percentage, $\Delta = \frac{N_{\phi}}{N}\cdot 100\%$. Rate adjustments are made by altering the degree of puncturing overhead.  Because puncturing operates on the factor in the denominator of the rate expression, the resulting reduction of this quantity increases the code rate.  Increasing the code rate results in fewer parity symbols transmitted per unit time, thus increasing throughput or reducing spectrum utilization for any desired target throughput.

For the block code we are implementing, the code rate after puncturing, $R_p$, is determined solely by the native (unpunctured) code rate and puncturing overhead, where $R_p = \frac{R}{1-\frac{\Delta}{100}}$.  Therefore, the amount of overhead required for any desired code rate is $\Delta = \left[ 1-\frac{R}{R_p} \right] \cdot{100\%}$. The codes outlined in \cite{Perrins1} are \textit{systematic} LDPC codes, wherein the source information sequence appears unaltered in the encoded sequence, but in general this is not a requirement because no distinction is made between information and parity symbols in our scheme and puncturing is performed after interleaving.  These codes exhibit sufficient regularity such that information symbols and parity symbols are equally protected and punctured symbols are therefore not limited to parity bits in our study.

For random puncturing, $N_{\phi}$ symbols are chosen at random from each codeword and removed as shown in Figure 3. These symbols are not sent over the transmission channel, but the puncturing pattern chosen must be known by the receiver in order to maintain the relative positions of the symbols in the received codeword during demodulation and decoding. It has been shown that the suitability for rate adjustment of an LDPC code ensemble by random puncturing is a function of the rate of the mother code and it's BP decoding threshold on the binary erasure channel (BEC) \cite{Mitchell1}. We explored this empirically through simulation of the ARTM0 code provided in \cite{Perrins1} on the additive white gaussian noise (AWGN) channel utilizing the transmitter and receiver models shown in the previous section. In the next section we present results confirming the robust decoding performance offered by randomly punctured LDPC codes on the AWGN channel with and without modulation. 

\section{NUMERICAL RESULTS}\label{sec:nums}

\noindent
In this section we present the results of our numerical simulations which were performed on the AWGN channel with and without CPM modulation. 

\begin{figure}[t]
\centering
\includegraphics[width = 5in]{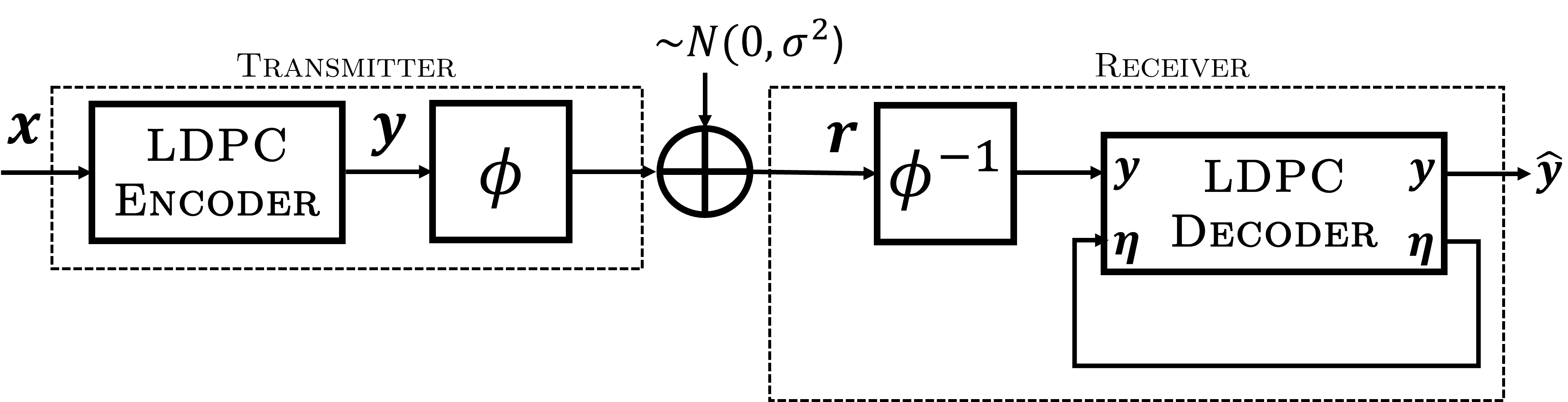}
    
\caption{Simplified transceiver model without CPM.}\label{LDPC2}
\end{figure}

\subsection{\MakeUppercase{Isolated Code}}
\begin{figure}[t]
\centering
\includegraphics[width = 5in]{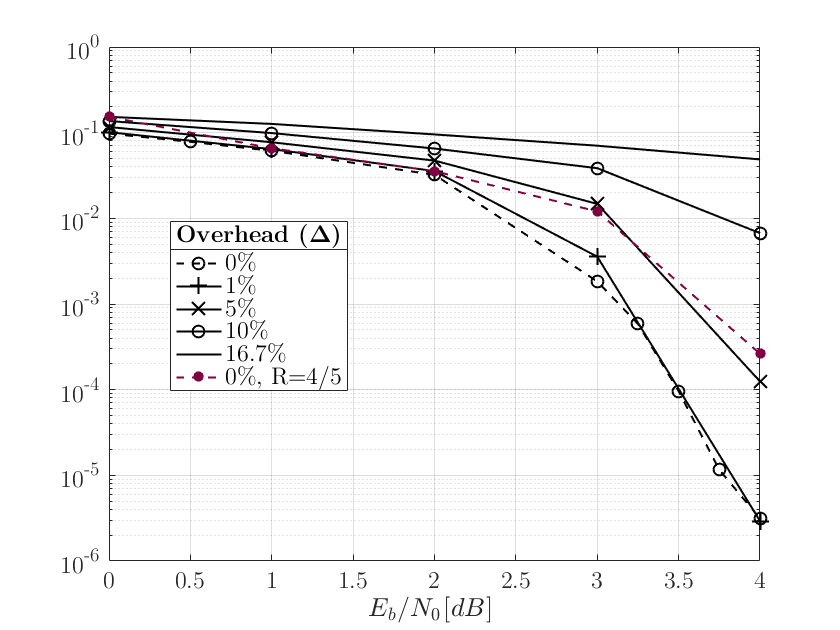}
    
\caption{BER performance for the $R=2/3$, $N=1024$ ARTM LDPC code with and without random puncturing.}\label{LDPC1}
\end{figure}

\noindent
In order to isolate the effect puncturing has on decoding performance of the LDPC code alone, we have removed CPM and associated support modules, resulting in the greatly simplified transceiver model shown in Fig. \ref{LDPC2}. With this model, puncturing is performed immediately before encoded information is sent over the communications channel, and depunctured after being received only once. Due to the elimination of the global iterative loop used to exchange information between the CPM and LDPC decoders (Fig. \ref{receiver}), the depunctured information directly enters the SPA LDPC decoder, where it is iteratively decoded to produce source information estimate $\hat{\textbf{y}}$. Simulated error correction performance on the AWGN channel for the $R=2/3$ ARTM0 code with blocklength $N=1024$ and puncturing overhead values of $\textbf{$\Delta$} = (0\%, 1\%, 5\%, 10\%, 16.7\% )$ is shown in Fig. \ref{LDPC1}. The value $\Delta=16.7\%$ is the percentage of puncturing necessary create a $R_p = 4/5$ code from a $R = 2/3$ mother code and thus represents the intersection between these two approaches. 

Without CPM, SPA decoding is strongly affected by random puncturing, with $~0.5$\,dB of coding gain lost at a BER of  $10^{-5}$ when moving from $\Delta = 0$ to $\Delta = 5$, which represents an effective rate increase from $R=2/3$ to $R_p=0.701$. Although small values of puncturing are well tolerated, such as shown for $\Delta = 1\%$, these do not provide a significant increase in code rate (\emph{e.g.}, $\Delta = 1\%$ increases $R=2/3$ to $R=0.673$). It is also observed that puncturing to the next available code rate of $R=4/5$ by setting $\Delta = 16.7\%$ results in poor performance, since the decoder cannot compensate for this large $\Delta$. Although it may happen that a code optimized for use with CPM is suboptimal for SPA decoding alone, it is worth nothing that a coding gain of over $7$\,dB was reported \cite{Perrins2} with the modulation scheme in place, outperforming all existing codes which were not optimized for the same modulation schemes. 

\subsection{\MakeUppercase{CPM Included}}
\begin{figure}[t]
\centering
\includegraphics[width = 5in]{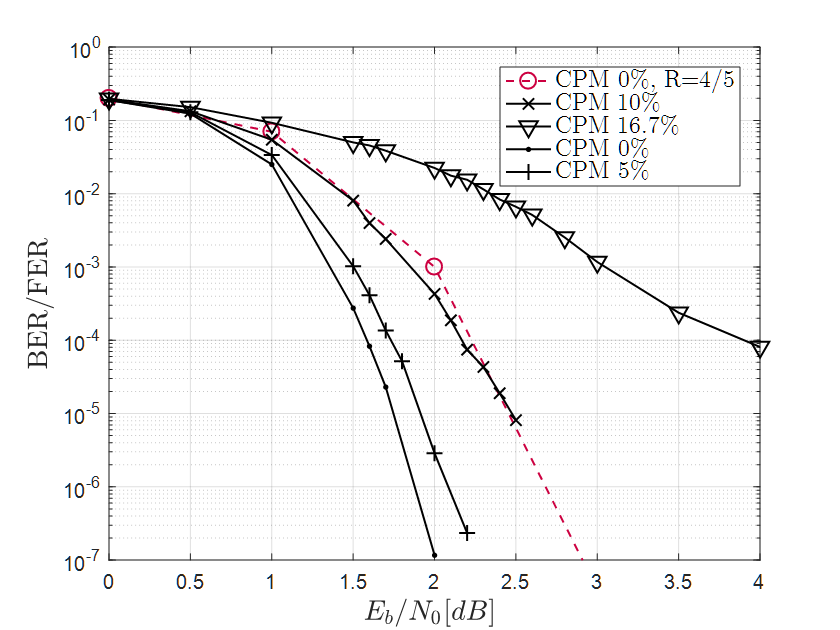}
    
\caption{BER performance with CPM for the punctured $R=2/3$ and unaltered $R=3/4$ ARTM LDPC codes with $N=1024$.}\label{CPM1}
\end{figure}

\noindent
With modulation included, as with the systems presented in Figs. \ref{transmitter} \& \ref{receiver}, a significant coding gain can be attributed to the global decoder-modulation loop alone.  As shown in Fig. \ref{CPM1}, a coding gain in excess of $2$ dB is observed between the systems with and without modulation even without considering puncturing ($\Delta = 0\%)$. This illustrates the chosen LDPC code's optimization for ARTM modulation and the ability of the SPA decoder to take advantage of information provided by the CPM decoder, working in concert by exchanging successively more accurate \textit{a priori} information in a doubly-iterative fashion. 

For the full system, puncturing has a much more gradual influence on the knee or \textit{waterfall} region of the resulting BER curve, with an \emph{error floor} only developing for overheads in excess of $10\%$. For example, with $\Delta = 5\%$, corresponding to an increase  of the code rate from $R=2/3=0.667$ to $R=0.702$, we observed a loss of approximately $0.5$\,dB at a BER of $10^{-4}$ in the LDPC only case (Fig.~\ref{LDPC1}), whereas in the CPM case we see that this is reduced to approximately $0.1$\,dB. At a bit error rate of $10^{-6}$, the loss remains less than $0.2$\,dB. At $\Delta = 10\%$, with $R_p = 0.74$, we see the punctured code still has reasonable performance (unlike no CPM). At this overhead, it performs similarly to the next highest available rate code of $R=4/5$ indicating that, at this point, it may be preferable to use the next code in the standard for higher rates rather than puncture the lower rate codes.  $R_p = R = 4/5$ is not achieved until $\Delta = 16.7\%$, at which point the performance dramatically degrades (like the case with no CPM). This occurs as a result of the LDPC decoder failing to converge satisfactorily. However, we conjecture that optimized puncturing patterns \cite{Mitchell2} may be able to reduce this gap. This is the subject of ongoing research.

Our simulation results provide evidence for augmentation of existing codes optimized for specific modulations with rate-compatibility by puncturing. In particular, our study demonstrates that small to moderate $\Delta$ can yield attractive trade-offs in rate vs. performance; however, it does not yet appear practical to achieve rates approaching or exceeding the next available ARTM LDPC code with random puncturing alone. From comparison with the non CPM results, the modulation feedback certainly appears to help. Furthermore, the less prominent error-floor features of the modulated version may indicate that more robust puncturing performance is possible for longer block lengths. In this regime, puncturing does not have as strong an impact, since most of the loss in performance due to puncturing comes from finite length effects of the SPA decoder and LDPC code (which itself improves with code length). 

\begin{figure}[t]
\centering
\includegraphics[width = 6.5in]{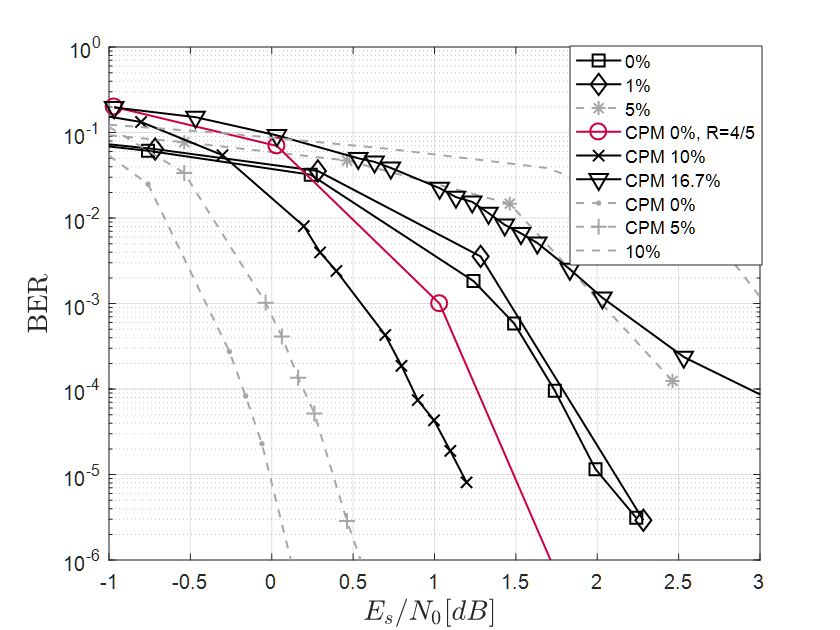}
    
\caption{Coding gain with CPM and puncturing included exceeds that of same $R=2/3$, $N=1024$ ARTM code without CPM.}\label{CPMvsLDPC}
\end{figure}

In order to better examine the region of interest around the $R=4/5$ target code, Fig. \ref{CPMvsLDPC} displays the relevant curves across the punctured $R=2/3$ (modulated and non-modulated) codes along with the unpunctured $R=4/5$ modulated ARTM code in terms of normalized symbol energy $E_s/N_0$. The gain between the punctured $\Delta = 10\%, R_p = 0.74$ modulated code and unpunctured native $R=4/5$ code is nearly $0.5$\,dB at a BER of $10^{-6}$, approximately the same as the loss in gain when going from the native $R=4/5$ code to punctured $\Delta = 16.7\%, R_p = 4/5$ code, highlighting that a window for further improvements is viable. We see that modulation has an even more prominent effect on performance than puncturing until the overhead reaches $\Delta \geq 10\%$, at which point the error floor dominates performance. This indicates that the puncturing has less of an impact on the CPM decoder than the LDPC decoder, perhaps due to the ASM marker being unaffected by puncturing under our scheme.  

Simulation results presented previously \cite{Perrins2} confirm formation of an error floor at a BER of $10^{-6}$, perhaps indicating an underlying weakness in that particular code that is exacerbated by puncturing with or without CPM decoding. At low values of overhead $\Delta \leq 5\%$, no error floor is yet apparent at $10^{-6}$ and thus error-floor effects seem to dominate only in high overhead regimes. This gives possible allowance for low overhead random puncturing to be realizable without any further considerations necessary. We also note that low-overhead puncturing where $\Delta \leq 10\%$ is well tolerated and performance is robust with only a gradual decrease in the slope of the BER curve within the waterfall region. Operation within the waterfall region is nevertheless practical and could serve as an adaptive mechanism when spectral bandwidth is limited and channel conditions are favorable. 

\subsection{\MakeUppercase{Hardware Considerations}}
\noindent
Random puncturing may be implemented by the insertion of a switch in the registers that serve as buffers to the inputs of each module in the signal chain. At random intervals chosen by the overhead parameters, block length, and system bus speed, the switch is placed in the high impedance position for one clock cycle. Depuncturing is accomplished by insertion of a random delay, at which point a zero-value is inserted into the buffer. Although this introduces additional latency, because all modules operate blockwise, it is possible to perform both puncturing and depuncturing in parallel. 

Thus when operating in a low-overhead regime as our results suggest, random puncturing offers robust rate-compatible performance with no changes to the submodules of Figs. \ref{transmitter} and \ref{receiver}.  Since no substantive hardware alterations are required, this allows for new operational modes at little expense in terms of latency and memory. However, it is worth noting that because the LDPC SPA decoder still operates on a depunctured sequence, SPA decoding computational complexity will remain largely unchanged.

\section{\MakeUppercase{CONCLUDING REMARKS}}\label{sec:conclusion}

\noindent
In this paper we have laid the groundwork for the construction of rate-compatible LDPC codes for IRIG-106 waveforms by the use of random puncturing. Our initial results show robust performance with codes designed for deployment with ARTM CPM modulation and open promising avenues for further improvement. Such systems would allow for the adaptive allocation of bandwidth in response to changing demand or channel conditions without requiring significant changes to the underlying encoding and decoding hardware and thus represent a low-complexity, low-cost implementation option. Furthermore, because puncturing is a type of erasure channel, it is easily modeled and those results are readily extended to other channel models, reducing design iteration schedules. 

By isolating the LDPC encoder and decoder, we were able to show the considerable coding gain afforded by the CPM-LDPC decoder loop, reinforcing that codes designed specifically for these modulations are better optimized than codes that do not consider modulation during design.  Though in neither case did we observe that random puncturing alone achieves the same performance as a code designed for that specific target rate, the gap is much more tractable with modulation, having a span of $0.5$\,dB, whereas without modulation the required overhead elicits an impractical error floor.  

In future work, other codes, including those of different rates and lengths, will be considered in order to better understand the interplay between the strength of the code and its resilience to puncturing under CPM. Modulation appears to be no impediment to the implementation of puncturing, with perturbations in the codeword's structure most strongly felt within the inner LDPC decoding loop of Fig.~\ref{receiver}.  Because the code we are using in our model is designed specifically to take advantage of information from the CPM decoder, further work should examine the selection of puncturing patterns which least disturb this feedback mechanism in order to preserve its portion of the total coding gain, which was observed to be in excess of $2$\,dB. 

We also plan to expand and generalize the initial work presented here by performing density evolution analysis on the BEC for these LDPC codes, thereby better determining their decoding threshold under random puncturing \cite{Mitchell1}. This knowledge will help inform the selection of nonrandom or quasi-random puncturing patterns which may bridge the gap between the punctured and native codes seen in our present analysis. It will also better describe the limits of the code design procedure and may offer new avenues for optimization.  

\section*{ACKNOWLEDGMENT}\label{sec:secAck}
\noindent
This material is based upon work supported by the National Science Foundation under Grant Nos. CNS-2148358 and HRD-1914635. 


\bibliographystyle{ieeestr}

\end{document}